\documentclass[a4paper,twocolumn,10pt]{article}
\usepackage{eurosis}
\usepackage{url}
\usepackage{natbib}

 \usepackage[pdftex]{graphicx}
\bibpunct[; ]{(}{)}{,}{a}{}{;}

\usepackage{subcaption}
\usepackage{float}
\usepackage{amsmath}
\usepackage{amssymb}

\usepackage{amsthm}

\newcommand*{\Resize}[2]{\resizebox{#1}{!}{$#2$}}%

\newtheorem{theo}{Theorem}
\begin{document}

\title{\uppercase{Non Cooperative Game Theoretic Approach for Residential Energy Management in Smart Grid}}

\author{Ilyes Naidji \\
Faculty of Mathematical, Physical and Natural Sciences,\\ University of Tunis-El Manar, Tunisia.\\
Email: ilyes.naidji@fst.utm.tn
\and
Moncef Ben Smida \\ LSA Lab, Tunisia Polytechnic School,\\ University of Carthage, Tunisia.\\
Email: moncef.bensmida@gmail.com
\and
Mohamed Khalgui\\
National Institute of Applied Sciences and Technology,\\ University of Carthage, Tunisia.\\
Email: khalgui.mohamed@gmail.com
\and
Abdelmalik Bachir \\ LESIA Lab, Computer Science Department,\\ Biskra University, Algeria.\\
Email: a.bachir@univ-biskra.dz
}
\date{}
\maketitle

\thispagestyle{empty}


\begin{abstract}
Demand side management (DSM) is one of the main functionalities of the smart grid as it allows the consumer to adjust its energy consumption for an efficient energy management.
Most of the existing DSM techniques aim at minimizing the energy cost while not considering the comfort of consumers. Therefore, maintaining a trade-off between these two conflicting objectives is still a challenging task.
 This paper proposes a novel DSM approach for residential consumers  based on a non-cooperative game theoretic approach, where each player is encouraged to reshape its electricity consumption pattern through the dynamic pricing policy applied by the smart grid operator. 
The players are guided to select the best strategy that consists of scheduling their electric appliances in order to  minimize the daily energy cost and  their discomfort level. The Nash Equilibrium of the energy management game is achieved using Non-Sorting Genetic Algorithm NSGA-II. 
Simulation results show the effectiveness of the distributed non cooperative game approach for the residential energy management problem where an appreciable energy cost reduction is reached while maintaining the discomfort in an acceptable level. 
\end{abstract}

\section{INTRODUCTION}
Along with the current transition of the power system from a centralized to a distributed architecture~\citep{mosbahi2016new}, a great attention is being paid to the power grid's capacity to maintain the balance between demand and supply~\citep{meskina2017multiagent, abidi2017multi}. 
Indeed, the increasing penetration of distributed renewable energy sources (RES) which have an intermittent nature  induces new challenges for the smart grid in terms of energy management, congestion, voltage and frequency variations, etc. In order to overcome these challenges, the demand side management (DSM) is increasingly exploited by smart grid operators to maintain the demand-supply equilibrium taking advantage of the demand flexibility.
DSM brings many solutions for consumers such as energy saving through the reduction of the electricity consumption and best use of electric appliances. Recent advances in information and communication technologies offer the opportunity for advanced DSM solutions, e.g., demand response, time of use, spinning reserve, etc.
In this study, we investigate the demand response (DR) solutions to reduce the energy cost for consumers while ensuring their comfort. The DR solutions consist of the short-term changes in the power consumption that could be made in response to the energy price variation. The dynamic pricing is designed to incite consumers to participate in the DSM by decreasing or increasing their power consumption. In addition, DR solutions do not only consist in reducing the power consumption but can modify the consumption pattern. 
DR is enabled through communication infrastructures~\citep{fadel2015survey}, allowing to decrease energy consumption during peak periods. It has been shown that DR can solve some existing problems in traditional power systems and enhance the reliability~\citep{safamehr2015cost,ma2013demand}.

In conventional  power systems, electricity prices do not change to solve reliability problems, thus consumers are not motivated to adjust their electricity consumption.
Smart grid technologies enable another pricing methods in restructured form. The prices are variable with respect to the demand and operating conditions, which involve the consumers participation in the power system operation.

Various methodologies have been proposed for the energy management of the smart grid using demand response. In~\citep{al2017advanced}, advanced demand response is proposed considering modular and deferrable loads with the objective of reducing the cost of consumed energy and peak consumed power.
In~\citep{vivekananthan2014demand, kaddah2014advanced}, direct load control (DLC) programs have been proposed. DLC allows the energy provider  to directly control (switch on/off) the electric appliances of the consumers with respect to their agreements (e.g., maximum number of interruptions, appropriate rewards, etc).
Simply shifting the power consumption of consumers at a peak time to off-peak times may cause consumer's discomfort. To model the discomfort of consumers, the difference between the desired load and the scheduled load is considered in~\citep{deng2014residential}.When shifting the power consumption pattern, the difference between the desired and scheduled time of loads can be considered to evaluate consumer's discomfort.
In~\citep{eksin2015demand}, a game theoretic approach was used to solve the demand response using the Bayesian Nash equilibrium with the objective of minimizing the peak-to-average ratio. However the comfort level of consumers was not addressed.
In~\citep{ning2017bi},  a coordinated optimization  is proposed where the concept of demand response potential (DRP) was introduced. However, the approach did not verify the consumer's comfort with concrete result.

Most of the aforementioned studies focus on the reduction of the energy cost and peak load to solve the energy management problem and do not sufficiently consider consumer behavior. In particular, the comfort level of consumers was not considered simultaneously with the energy cost in the demand side studies. Even when jointly considered with the energy cost like in \citep{yang2013game,kim2013bidirectional}, they are subsequently referred as a total cost. Such consideration may affects the result of the energy management system and may give a biased solution sometimes for the energy cost and sometimes for consumers comfort.

In this respect, we propose a multiobjective demand response game for the optimal scheduling of the electric appliances in smart homes. Non cooperative energy management game is designed to model the behavior of consumers. The proposed energy management game incorporates the demand side in the supply management using the dynamic pricing policy to minimize daily energy cost while reducing the discomfort level of consumers with a multiobjective approach.
 
The contributions of this paper can be summarized as follows:
\begin{itemize}
\item{We propose a non cooperative energy management game which guarantees the fairness among non-cooperative consumers and apply the NSGA-II algorithm to find the Nash equilibrium of the game in accordance to the two conflicting objectives that are energy cost and consumer's comfort.}
\item{We model the rational behavior of the consumers that reduce their energy cost while seeking their comfort.}
\item{We address the concept of discomfort level of consumers which is based on the difference between the desired and scheduled time of the electric appliances.}
\end{itemize}
This paper is organized as follows: Section~\ref{sec:system_model} gives the smart grid model used in this paper. Section~\ref{sec:game_formulation} explains the  concept of game theory for the energy management problem. Section~\ref{sec:problem-form} gives the formulation of the proposed  energy management game. Section~\ref{sec:num_results} gives the numerical results. Finally, Section~\ref{sec:conclusion} concludes this paper.

\section{\uppercase{System Model}}
\label{sec:system_model}
\begin{figure}[ht]
\centering
\includegraphics[width=0.43\textwidth]{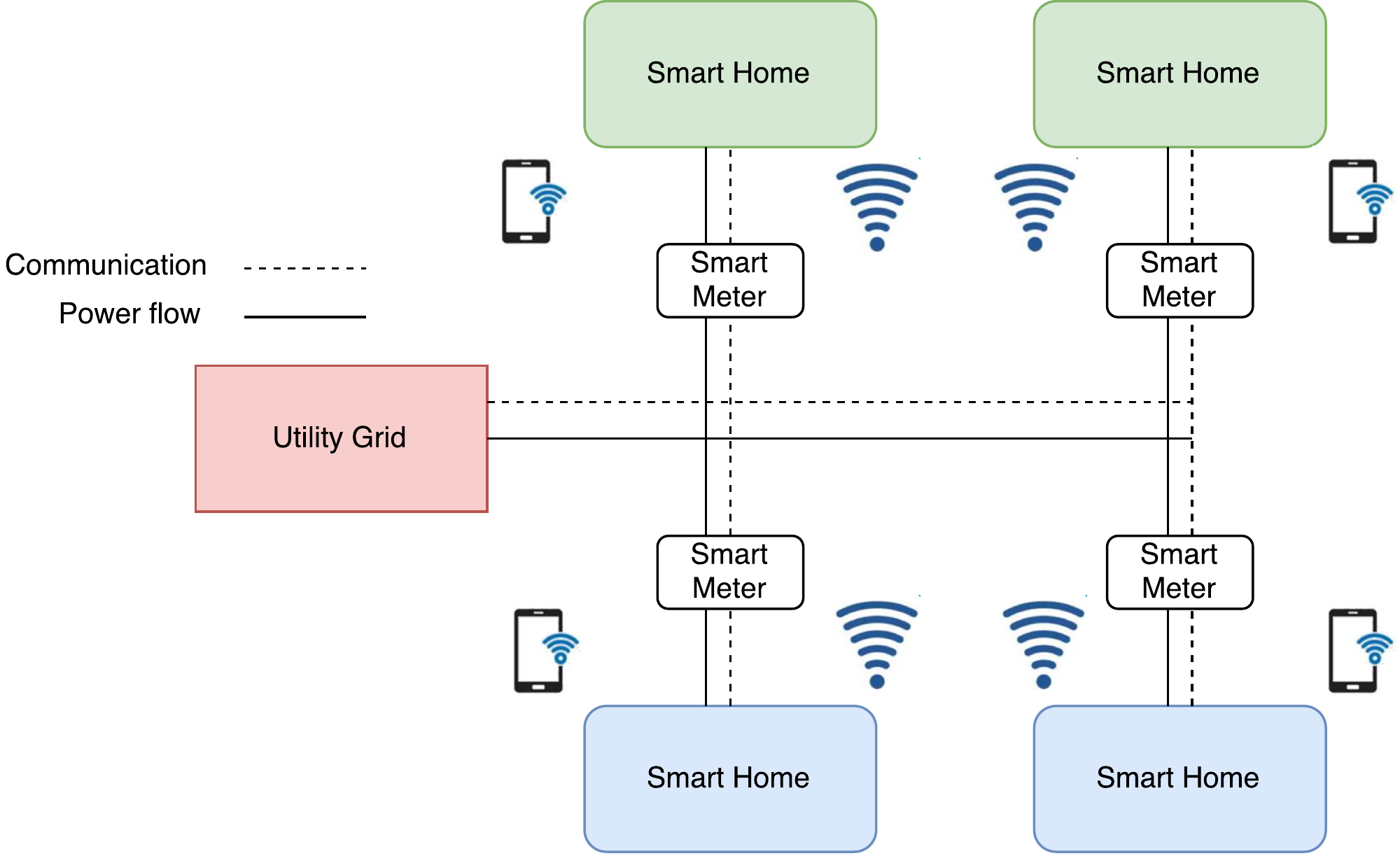}
\caption{Distributed EMS Architecture.}
\label{arch}
\end{figure}

This section describes the smart grid system used in this study and gives the energy management system (EMS) architecture. 
Consider a smart distribution grid with one electricity provider that supplies energy to a set of $n$ consumers, i.e., smart homes which are controlled by a Multi-Agent System (MAS).  
Each smart home is equipped with a smart meter that is connected to the electric appliances via wire connection e.g., PLC (Power Line Communication) or wireless connection, e.g., ZigBee, etc.
Furthermore, the smart meter integrates an agent that has computational intelligence capabilities.
Each agent collects the planned tasks for the current day and power consumption profiles of the electric appliances from the smart meter. 
Consumers communicate with the smart meter through wireless communication, e.g., smart phones or tablets, to indicate the preferred conditions of their electric appliances related to their comfort, e.g., the preferred temperature in the rooms, desired time to charge the electric vehicle, etc.
Fig.~\ref{arch} shows the architecture of the proposed EMS.

In the proposed EMS, each consumer is characterized by its planned daily tasks $TS_j$. 
 Let $TS = \{TS_1, TS_2, ..., TS_k\}$ be the set of daily tasks to be executed, each task can be characterized by two vectors~\citep{salinas2013multi}:
\begin{equation}
X_j = \bigg[x_j^1, x_j^2, ..., x_j^t\bigg]
\end{equation}
\begin{equation}
Y_j= \bigg[ P_j \quad D_j \quad ST_j \quad FT_j \quad STP_j \quad FTP_j \bigg]
\end{equation}
where 
\begin{itemize}
\item{${X}_{j}$ is the power consumption profile of the electric appliance that executes the task $j$ where ${x}_j^t$ is the power consumption of task $j$ at time $t$}.
\item{${Y}_{j}$ is the vector which characterizes the task $j$.}
\item{$P_j= \underset{t=1}{\overset{T}{\sum}} x_j^t$ is the energy demand of the task $j$ where the time horizon $T=24$.}
\item{$D_j$ is the duration of the task $j$.}
\item{$[ST_j$, $FT_j]$ are the earliest start time and finishing time to run the task $j$ that define its admitted interval of execution.}
\item{$[STP_j, FTP_j]$ is the  time preferred window of the consumer to run the task $j$.}

\end{itemize}

\subsection{Pricing method}
Let $P_{ij}^{t}$ is it the energy consumption of consumer $i$ for task $j$ at time $t$. The total energy consumption of all consumers $(i=1,..., n)$ at time slot $t$ is defined as follows:
\begin{equation}
l^t= \overset{n}{\underset{i=1}{\sum}}\overset{k}{\underset{j=1}{\sum}}P_{ij}^t
\end{equation}
Consider $C_u^t(l^t)$ the cost function of utility grid at time slot $t$. 
 The quadratic cost function is usually used in literature~\citep{deng2014residential}, i.e, 
 \begin{equation}
 C_u^t(l^t)= a^t.(l^t)^2+ b^t.l^t +c^t 
 \end{equation}
 where  the quadratic costs coefficients $(a^t, b^t, c^t)$ are time varying.
 Real-time pricing is used in this study where the price value is time varying. The price value depends on time of use (TOU) and total energy consumption. Based on this pricing model, the energy cost of consumers at time slot $t$ is defined as in~\citep{deng2014residential} by:
 \begin{equation}
C_c^t= c_r^t(l^t). \overset{k}{\underset{j=1}{\sum}}P_j^t
\end{equation}
where $c_r^t(l^t)$ is the real-time price of energy. We assume that the smart grid operator adopts adequate pricing method that takes into consideration the energy consumption in time and level. In this study, we consider also consumers within distributed energy resources (DERs) facility, i.e., prosumers that can generate energy.
The revenue of the prosumer $R_p^t$ is calculated as follows:
 \begin{equation}
R_p^t=  \overset{T}{\underset{t=1}{\sum}}r.P_p^t
\end{equation}
where $P_p^t$ is the power generated from the prosumer at time $t$ and $r$ is the revenue coefficient.
\subsection{Consumer discomfort level}
To measure the discomfort of consumers caused by shifting their consumption pattern, we introduce a discomfort cost as a quadratic function of the gap between the desired and the scheduled time of the electric appliances. 
 
We define a time shift parameter $\Delta_j$ that models the gap between the scheduled   and  preferred time of the task $j$. Let $t_j$ denotes the start time of the task $j$.
The time shift parameter  $\Delta_j$ is calculated as follows:
  \begin{equation}\label{timeshift}
 \Delta_j  = 
  \begin{cases}
 	0, & \text{if }     \Resize{4cm}{t_j \ge STP_j \land t_j + D_j \le FTP_j,}
  \\  \Resize{1.4cm}{STP_j - t_j} , & \text{else if }  \Resize{4cm}{t_j \le STP_j  \land t_j + D_j \le FTP_j,}
    \\  \Resize{2cm}{(t_j + D_j) - FTP_j}   , & \text{else if }  \Resize{4cm}{t_j \ge STP_j  \land  t_j + D_j \ge FTP_j,}
  \end{cases}
   \end{equation}
Here, for each task $j$, the concept of time shift can be valid only in the admitted interval of execution $[ST_j, FT_j]$ of the task $j$.
The discomfort cost can be modeled with a quadratic cost function \citep{samadi2012advanced} as follows:
 \begin{equation}
C_{ \Delta_j} = \overset{k}{\underset{j =1}{\sum}} \alpha ( \Delta_j)^2 + \beta  \Delta_j + \delta 
\end{equation}
 where $\alpha$, $\beta$ and $\delta$ are the quadratic cost coefficients. Here, as more as the time shift parameter increases, i.e., the electric appliance is scheduled out of its preferred time, the discomfort cost increases.

\section{\uppercase{Mathematical Game Theory Formulation}}
\label{sec:game_formulation}
This section gives the mathematical game theory formulation of the energy management problem involving the MAS.

In the energy management problem, the consumers do not collaborate, for example, when a smart meter shows the real-time electricity price in the smart grid, the consumer reduces or increases its electricity consumption without asking neighbors whether they reduce their consumption or not at a certain time. Game theory is able to model the competitive behavior of the consumers.
In the proposed energy management architecture, each consumer, i.e., agent is a player of the non-cooperative game. 
The energy management game can be defined by the following 3-tuple:
\begin{equation}
G= \{ N, S, J\}
\end{equation}
where $N$ is the set of players with $|N| = n$. $S$ is the strategy space of players where 
\begin{equation}
S = S_1 \times S_2 \times ... \times S_n  
\end{equation}
$J = S \rightarrow \mathbb{R} $ is the vector of cost functions of players $i=1, 2, ..., n$ which is defined as 
\begin{equation}
J(s) =  [J_1(s), J_2(s), ..., J_n(s)] \quad s\in S
\end{equation}
where the vector of strategies $s= (s_1, s_2, ..., s_n) \in S$  is called a strategy profile.
Let   $\{a_i\}$ be the set of actions of the player $i$. Each action $a_i$ represents the total energy consumption of player $i$ over the time slot $t$. 
For player $i$, the set of the selected actions consists of the energy consumption pattern for the time horizon $T=24$, i.e., one day.  
The strategy $s_i$ of the player $i$ can be regarded as a rule for choosing its actions. The cost function of the player $i$ is given by
\begin{equation}
J_i(s)= J_i \{s_i^{*}, \overline{s_{i}^{*}}\}
\end{equation}
where the cost function $J_i$ depends on the strategy $s_i^{*} \in S_i$ selected by the player $i$ and on the strategy profile $\overline{s_{i}^{*}}$ of the other players.
Solving the energy management game consists of finding the Nash Equilibrium for each player in the non cooperative game. The vector $s^{*} = (s_1^{*}, s_2^{*}, ... , s_n^{*})$ is a Nash equilibrium for the energy management game $G= \{ N, S, J\}$  if the following constraint is valid:
\begin{equation}
\forall i \in N, \forall s_i \in S_i, \quad J_i(s_i^{*},\overline{s_{i}^{*}}) \le J_i(s_i,\overline{s_i^{*}})
\end{equation}

\section{\uppercase{Problem Formulation}}
\label{sec:problem-form}
This section describes the energy management game and the considered game players.
\subsection{Energy management game players}
The proposed energy management game consists of two types of players that are: consumer denoted by (c-player) and prosumer denoted by (p-player). C-player represents the flexible consumers that can adjust their consumption pattern. 
P-player represents the consumers within DERs facility. Each player has its own objective functions, as illustrated in the following.
\subsubsection{C-player}
The c-player can adjust its consumption pattern through managing its smart electric appliances. The first objective of the c-player is to reduce the energy cost as follows:
\begin{equation}
\min \quad J_1 = \overset{T}{\underset{t=1}{\sum}} C_c^t
\end{equation}

 The second objective of the c-player is to minimize its discomfort level by
\begin{equation}
\min \quad J_2 =  \overset{k}{\underset{j =1}{\sum}} \alpha (\Delta_j)^2 + \beta \Delta_j + \delta
\end{equation}
Hence, the cost function of the c-player is given by
\begin{equation}
J_{c}(s) = (J_1, J_2)
\end{equation}
\subsubsection{P-player}

The p-player can adjust its consumption pattern and manage its DERs. The first objective of the p-player is to minimize its energy cost and maximize its revenue as follows:
\begin{equation}
\min \quad  J_3 = \overset{T}{\underset{t=1}{\sum}} C^t_{p} -  R^t_{p} 
\end{equation}
where $ C^t_{p}$, $R^t_{p}$  are the energy cost and revenue of p-player at time slot $t$, respectively.

 The second objective of the p-player is to minimize its discomfort level by
\begin{equation}
\min \quad J_4 =  \overset{k}{\underset{j =1}{\sum}} \alpha (\Delta_j)^2 + \beta \Delta_j + \delta
\end{equation}
Thus, the cost function of the p-player is given by
\begin{equation}
 J_{p}(s) = (J_3, J_4)
\end{equation}

\subsection{Constraints}
The proposed energy management game is subject to the following constraints.
\subsubsection{Time constraints}
Each task $j$ of duration $D_j$ must be executed exactly once between its earliest start time $ST_j$ and finishing time $FT_j$. The start time $t_j$ of task $j$ satisfies the following constraint:
\begin{equation}
ST_j \le t_j \le FT_j - D_j 
\end{equation}
\subsubsection{Energy balance} The energy generated by the utility grid and p-player must be equal to the total energy consumed by c-player and p-player satisfying:
\begin{equation}
E^t_{u} +  E^{t}_{p}  - E^{t}_{d}  =  0   
\end{equation}
where $E^t_{u}$ is the energy produced by the utility grid, $E^{t}_{p}$ is the energy produced by the p-player and $E^{t}_{d}$ is the total energy demand of all consumers, i.e., c-players and p-players. 
\subsubsection{Nash equilibrium} 
The existence of the Nash equilibrium is proved by the following theorem \citep{nash1951non}:

\begin{theo}
Every game with a finite number of players that can choose from finitely number of strategies has at least one Nash equilibrium. 
\end{theo}

In the proposed non cooperative game, each player is assumed to be rational, i.e., the player aims to minimize its cost function by considering the best strategy. Therefore, each player chooses the load profile that represents its best strategy. 
\begin{theo}
The combination of best strategies and their corresponding cost functions constitutes the dominant strategy which is the Nash equilibrium for the energy management game \citep{fudenberg1991game}. 
\end{theo}
\subsection{NSGA-II scheduling}

NSGA-II algorithm is used to find the Nash equilibrium of the energy management game. A detailed description of the algorithm can be found in~\cite{deb2002fast}.  
 Each player, i.e., agent gets the information about time-differentiated electricity price from the smart grid operator to adjust the scheduling of its daily tasks. After that, 
 the agent applies the NSGA-II algorithm to solve the multiobjective optimization problem with the objective to minimize the daily energy cost and consumer's discomfort.
Hence, the NSGA-II algorithm specifies the best strategy of each player in accordance to the minimization of two fitness functions. 
For the c-player, the strategy is chosen by
\begin{equation}
s_i^{*c-best} = \arg \min \big[J_{c}(s)\big]
\end{equation}
For the p-player, the strategy chosen by
\begin{equation}
s_i^{*p-best} = \arg \min \big[J_{p}(s)\big]
\end{equation}
The NSGA-II scheduling solution gives the dominant strategy for each player which is the best strategy. The combination of these best strategies constitutes the dominant strategy which is the Nash Equilibrium of the energy management game~\citep{fudenberg1991game}.
\section{\uppercase{Numerical Results}}
\label{sec:num_results}
This section illustrates the system under study and gives the numerical results of the proposed multiobjective energy management game.

For testing the proposed method, the EMS is developed in MATLAB software. 
The system under study consists of a smart grid with $|N|=30$ consumers. The system includes several  c-players and p-players. Each player performs its daily tasks, at least 8 tasks selected randomly from Table~\ref{tab1}. 
\begin{table}[]
\centering
\caption{Task's characteristics.}
\label{tab1}
\resizebox{0.475\textwidth}{!}{
\begin{tabular}{lllll}
\hline
\multicolumn{1}{|l|}{Task}                                                       & \multicolumn{1}{l|}{\begin{tabular}[c]{@{}l@{}}Power \\ (kW)/h\end{tabular}} & \multicolumn{1}{l|}{\begin{tabular}[c]{@{}l@{}}ST\_j\\ (hour)\end{tabular}} & \multicolumn{1}{l|}{\begin{tabular}[c]{@{}l@{}}FT\_j\\ (hour)\end{tabular}} & \multicolumn{1}{l|}{\begin{tabular}[c]{@{}l@{}}Duration\\ (hour)\end{tabular}} \\ \hline
\multicolumn{1}{|l|}{Washing machine}                                            & \multicolumn{1}{l|}{1}                                                       & \multicolumn{1}{l|}{6}                                                      & \multicolumn{1}{l|}{24}                                                     & \multicolumn{1}{l|}{2}                                                         \\ \hline
\multicolumn{1}{|l|}{Laptop}                                                     & \multicolumn{1}{l|}{0.1}                                                     & \multicolumn{1}{l|}{18}                                                     & \multicolumn{1}{l|}{24}                                                     & \multicolumn{1}{l|}{6}                                                         \\ \hline
\multicolumn{1}{|l|}{Desktop}                                                    & \multicolumn{1}{l|}{0.3}                                                     & \multicolumn{1}{l|}{18}                                                     & \multicolumn{1}{l|}{24}                                                     & \multicolumn{1}{l|}{3}                                                         \\ \hline
\multicolumn{1}{|l|}{\begin{tabular}[c]{@{}l@{}}Air\\ conditionner\end{tabular}} & \multicolumn{1}{l|}{1.5}                                                     & \multicolumn{1}{l|}{10}                                                     & \multicolumn{1}{l|}{19}                                                     & \multicolumn{1}{l|}{1}                                                         \\ \hline
\multicolumn{1}{|l|}{\begin{tabular}[c]{@{}l@{}}Dish\\ Washer\end{tabular}}      & \multicolumn{1}{l|}{1}                                                       & \multicolumn{1}{l|}{7}                                                      & \multicolumn{1}{l|}{19}                                                     & \multicolumn{1}{l|}{3}                                                         \\ \hline
\multicolumn{1}{|l|}{Fridge}                                                     & \multicolumn{1}{l|}{0.3}                                                     & \multicolumn{1}{l|}{0}                                                      & \multicolumn{1}{l|}{24}                                                     & \multicolumn{1}{l|}{24}                                                        \\ \hline
\multicolumn{1}{|l|}{\begin{tabular}[c]{@{}l@{}}Electrical\\ Car\end{tabular}}   & \multicolumn{1}{l|}{3.5}                                                     & \multicolumn{1}{l|}{18}                                                     & \multicolumn{1}{l|}{8}                                                      & \multicolumn{1}{l|}{3}                                                         \\ \hline
\multicolumn{1}{|l|}{Boiler}                                                     & \multicolumn{1}{l|}{0.8}                                                     & \multicolumn{1}{l|}{15}                                                     & \multicolumn{1}{l|}{22}                                                     & \multicolumn{1}{l|}{2}                                                         \\ \hline
\multicolumn{1}{|l|}{Iron}                                                       & \multicolumn{1}{l|}{1.2}                                                     & \multicolumn{1}{l|}{10}                                                     & \multicolumn{1}{l|}{22}                                                     & \multicolumn{1}{l|}{1}                                                         \\ \hline
\multicolumn{1}{|l|}{\begin{tabular}[c]{@{}l@{}}Cooker\\ Microwave\end{tabular}} & \multicolumn{1}{l|}{1.7}                                                     & \multicolumn{1}{l|}{6}                                                      & \multicolumn{1}{l|}{9}                                                      & \multicolumn{1}{l|}{1}                                                         \\ \hline
\multicolumn{1}{|l|}{Spin Dryer}                                                 & \multicolumn{1}{l|}{2.9}                                                     & \multicolumn{1}{l|}{13}                                                     & \multicolumn{1}{l|}{18}                                                     & \multicolumn{1}{l|}{1}                                                         \\ \hline
\multicolumn{1}{|l|}{Television}                                                 & \multicolumn{1}{l|}{0.6}                                                     & \multicolumn{1}{l|}{19}                                                     & \multicolumn{1}{l|}{24}                                                     & \multicolumn{1}{l|}{3}                                                         \\ \hline
\multicolumn{1}{|l|}{Cooker Oven}                                                & \multicolumn{1}{l|}{5}                                                       & \multicolumn{1}{l|}{18}                                                     & \multicolumn{1}{l|}{19}                                                     & \multicolumn{1}{l|}{0.5}                                                       \\ \hline
\multicolumn{1}{|l|}{Cooker Hob}                                                 & \multicolumn{1}{l|}{3}                                                       & \multicolumn{1}{l|}{8}                                                      & \multicolumn{1}{l|}{9}                                                      & \multicolumn{1}{l|}{0.5}                                                       \\ \hline
                                                                                 &                                                                              &                                                                             &                                                                             &                                                                               
\end{tabular}
}
\end{table}
To obtain the Nash equilibrium of the energy management game, our proposed multiobjective formulation was submitted to NSGA-II algorithm. 

\begin{figure}[H]
  \begin{subfigure}[b]{0.475\textwidth}
    \includegraphics[width=\textwidth]{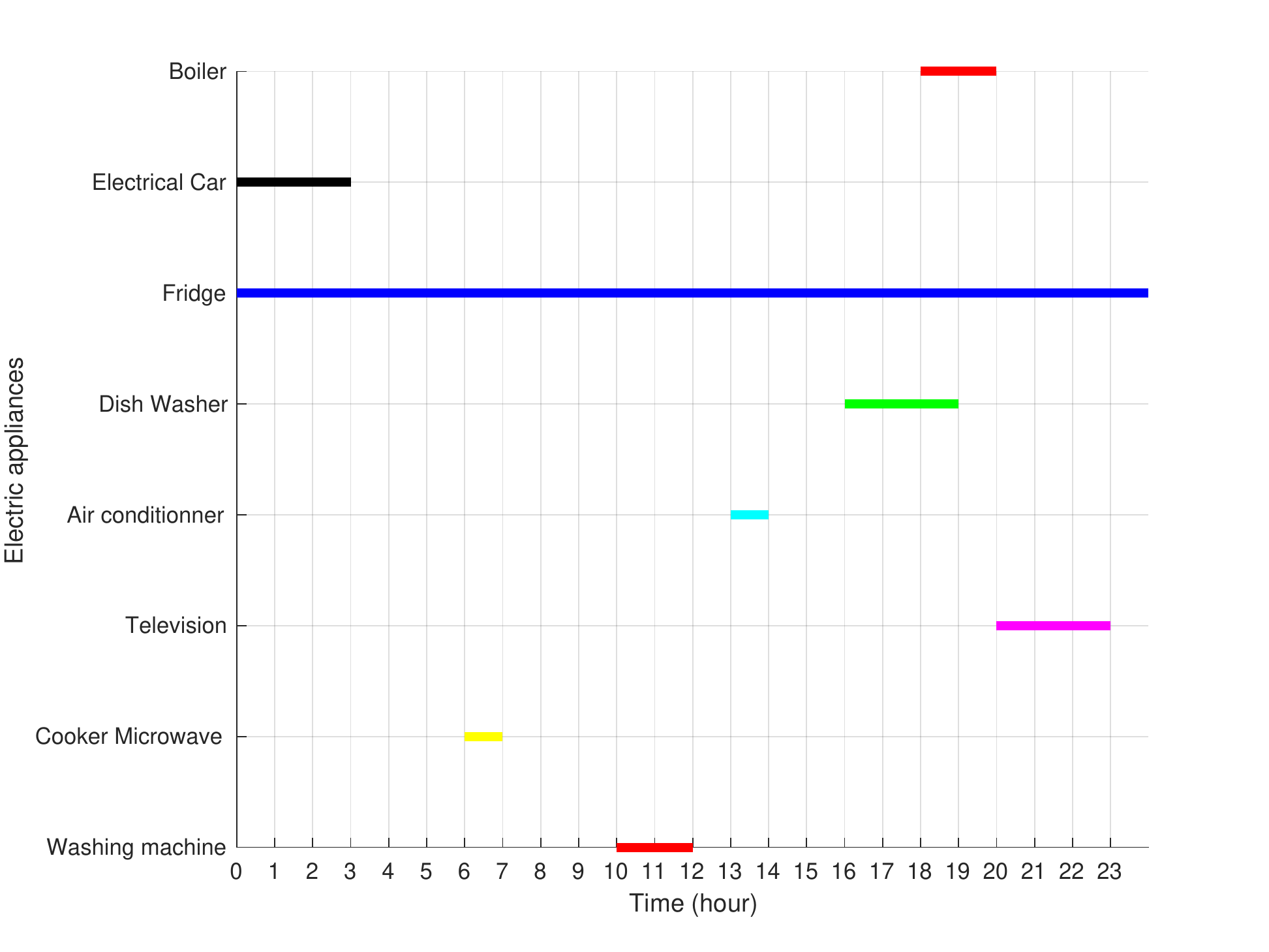}
    \caption{C-player.}
    \label{c-c2}
  \end{subfigure} 
   \begin{subfigure}[b]{0.475\textwidth}
    \includegraphics[width=\textwidth]{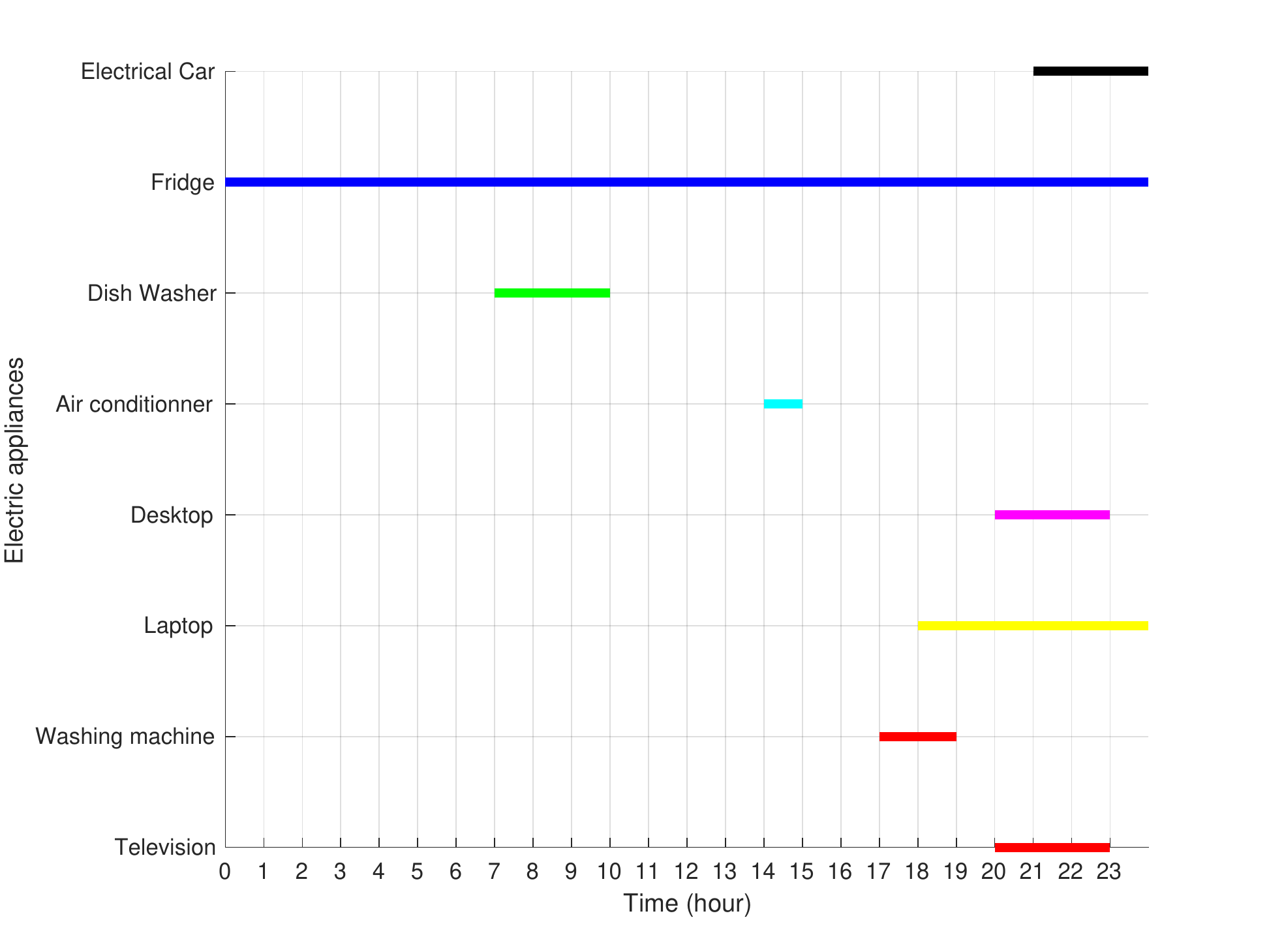}
    \caption{P-player.}
    \label{c-p2}
  \end{subfigure} 
  \caption{Tasks scheduling.}
   \label{tasks}
  \end{figure}

In order to incentivize the consumer to run some of its electrical appliance at particular time-slots of the day, the smart grid operator uses time-variable tariff rates. 
Fig.~\ref{tasks} summarizes the set of considered electricity appliances and highlights the starting and finishing time  of these appliances that can be scheduled in the admitted time window. The first subfigure shows the appliances of a typical c-player and the second subfigure shows the appliances of a typical p-player. 

To evaluate the effectiveness of the proposed scheduling solution, three scenarios are investigated. The first is the reference scenario (\lq Ref-sce\rq), where each task will be executed at its preferred time interval and will not be executed in early or later time. In the litterature, such scenario is called  the welfare maximization as in \citep{li2011optimal}.  
 The second is the cost effective scenario (\lq Cost-sce\rq) as in \citep{al2017advanced}, where the tasks are executed with the objective to minimize the daily energy cost without taking into account the discomfort of consumers.
The third scenario (\lq Cost-discomfort-sce\rq) refers to the proposed multiobjective scheduling solution which consists of the minimization of the daily energy cost and discomfort level of consumers.

\begin{figure}[H]
    \begin{subfigure}[b]{0.475\textwidth}
    \includegraphics[width=\textwidth]{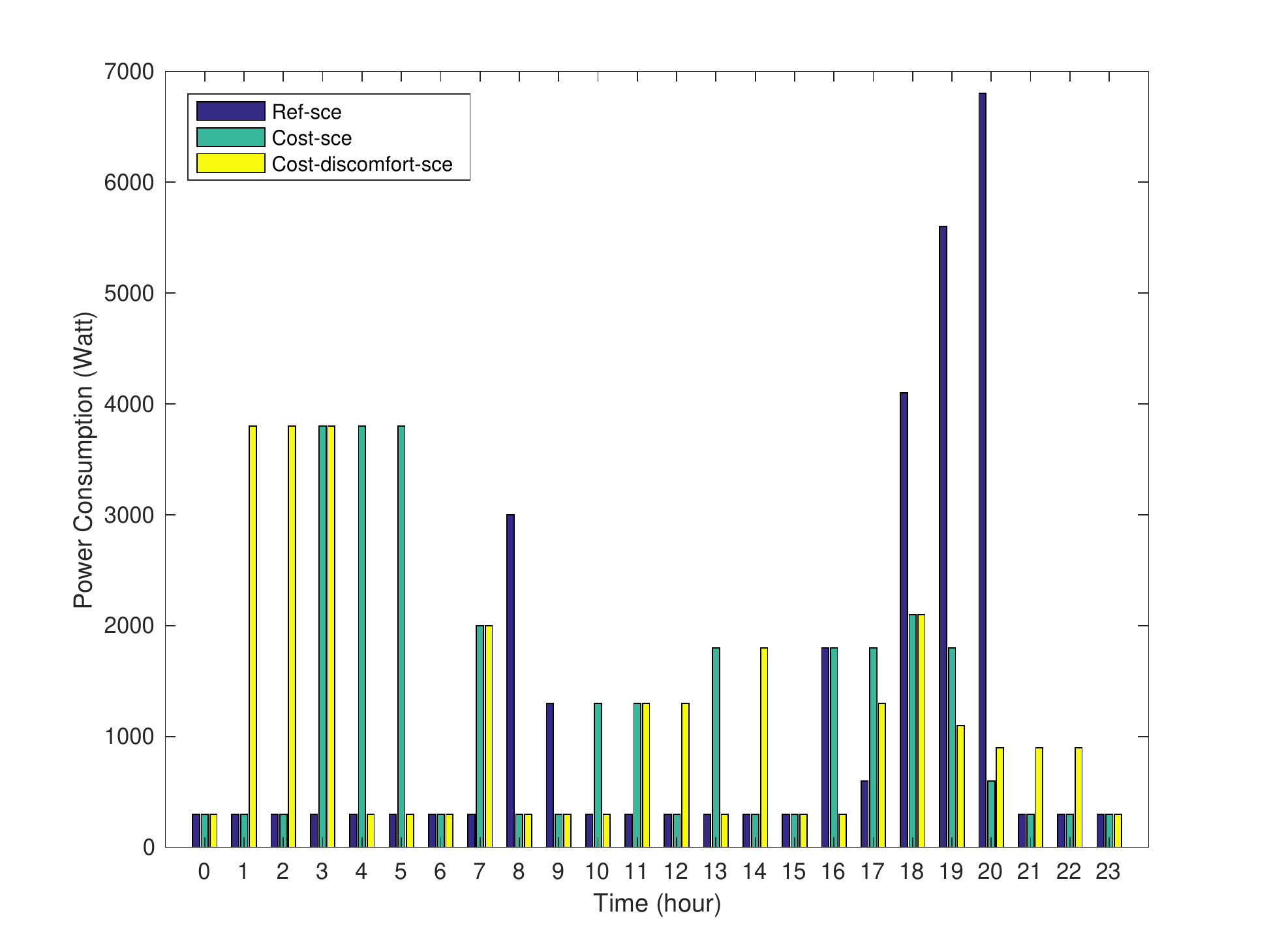}
    \caption{C-player.}
    \label{sci2}
  \end{subfigure}
   \begin{subfigure}[b]{0.475\textwidth}
    \includegraphics[width=\textwidth]{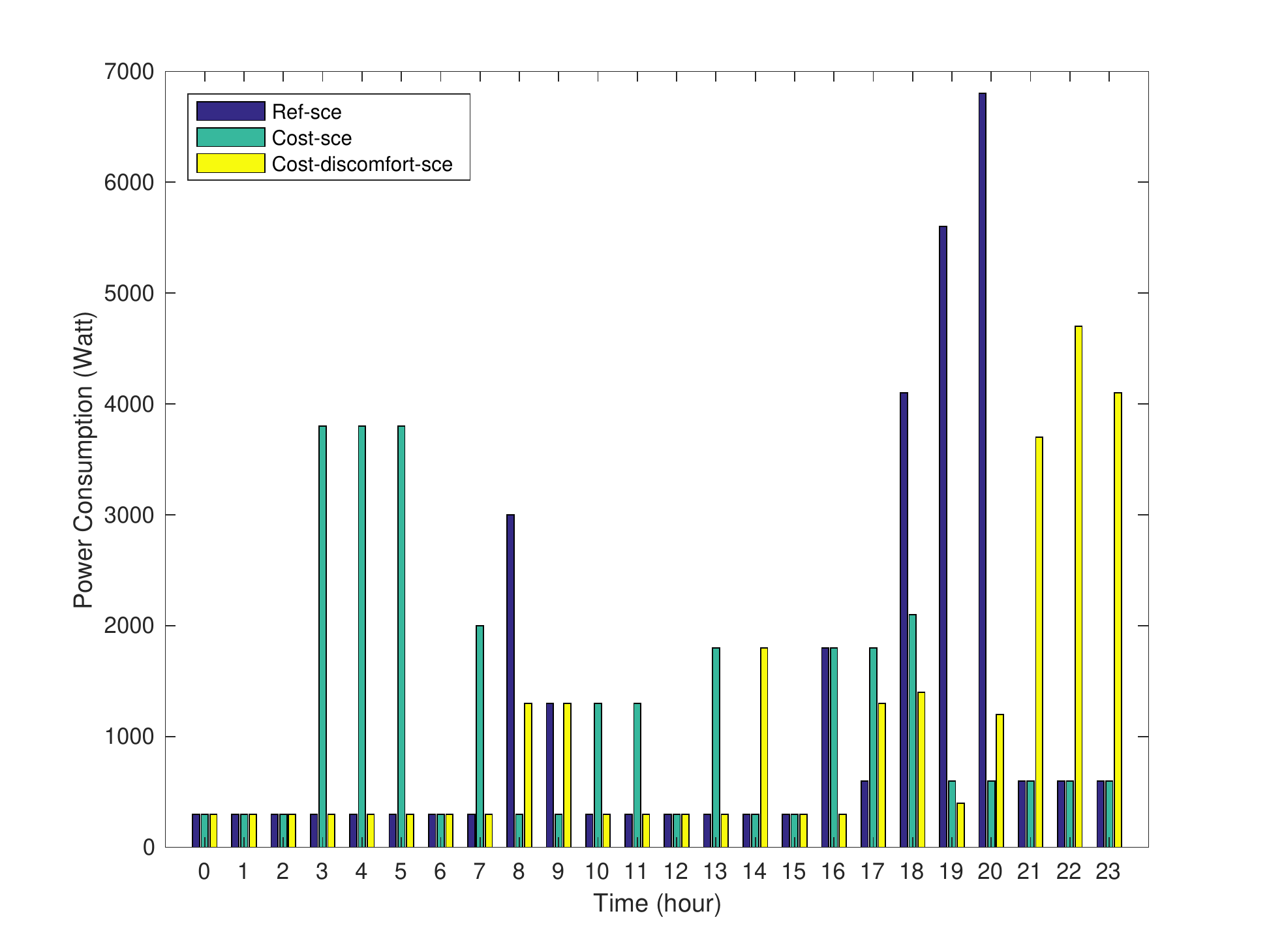}
    \caption{P-player.}
    \label{sci3}
  \end{subfigure}
      \caption{Power consumption.}
      \label{consumption}
  \end{figure}

\begin{figure}[H]
\centering
\includegraphics[width=0.475\textwidth]{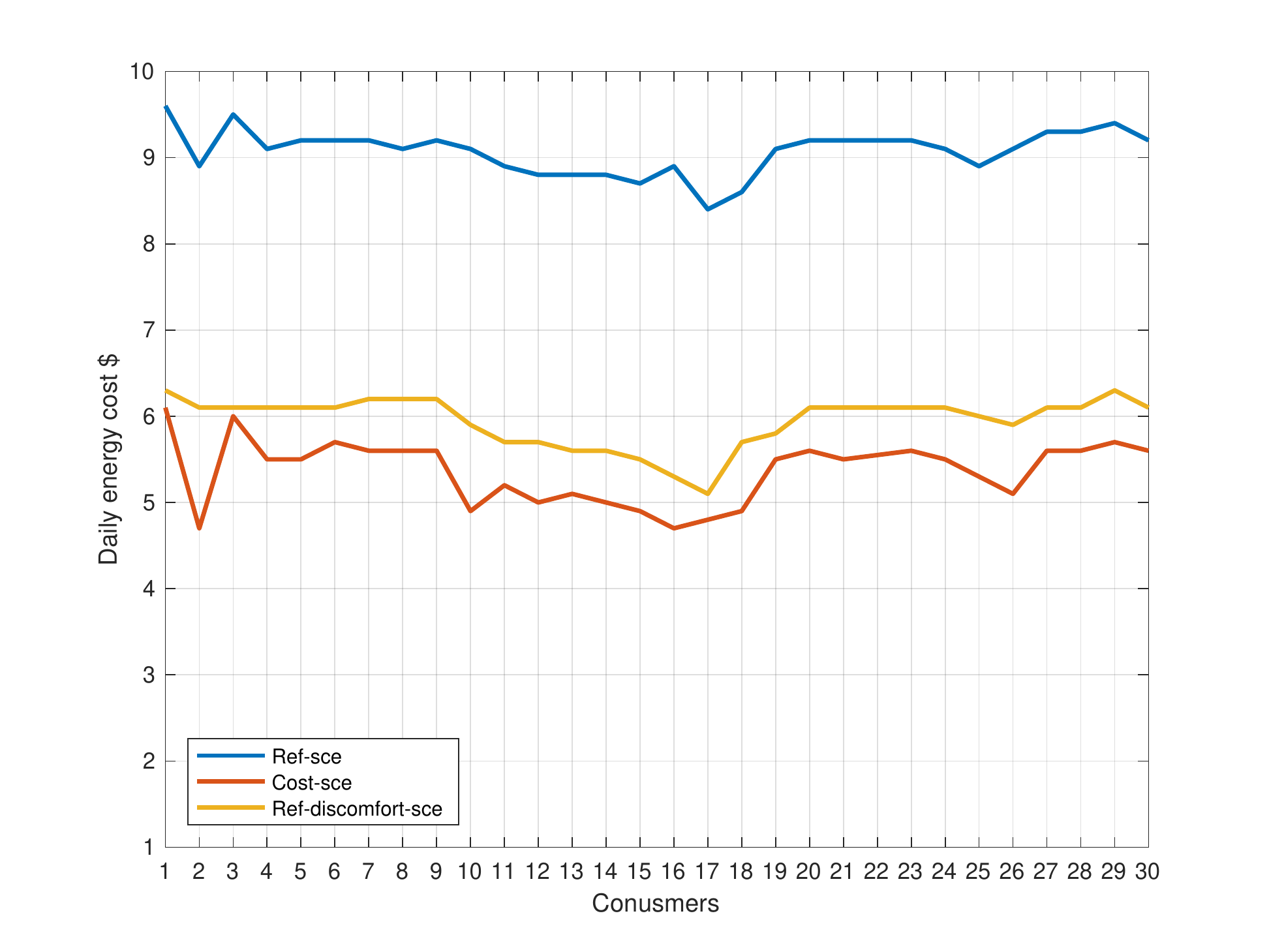}
\caption{Daily energy cost.}
\label{daily-cost}
\end{figure}

\begin{figure}[H]
\centering
\includegraphics[width=0.475\textwidth]{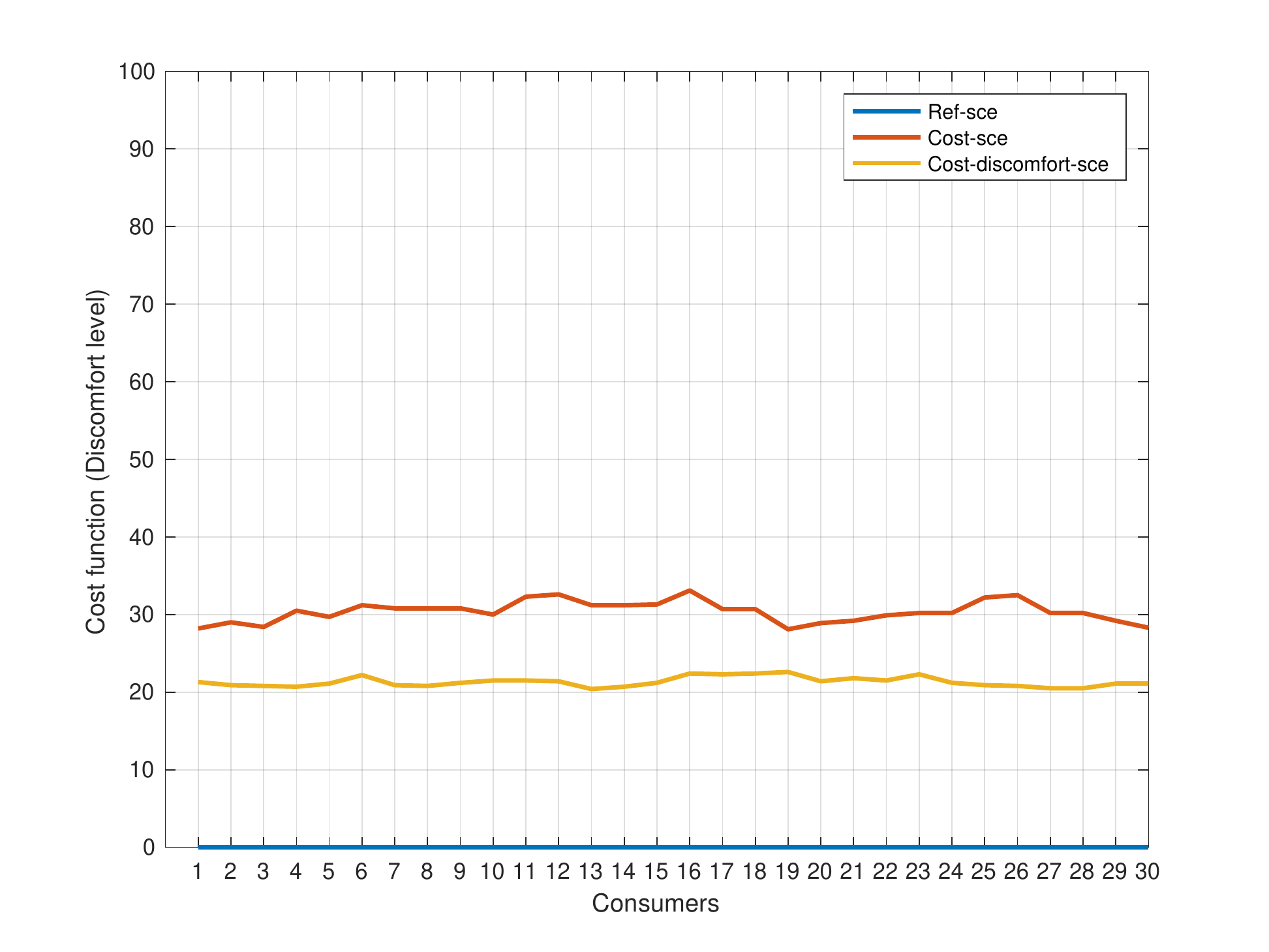}
\caption{Discomfort level.}
\label{payoff}
\end{figure}

Fig. \ref{consumption} shows the total power consumption for a typical c-player and a typical p-player for the considered scenarios, while Figs. \ref{daily-cost} and \ref{payoff} show the results of the considered scenarios for all the consumers in terms of energy cost and discomfort level.
  It can be seen that p-player has more comfort than c-player taking the advantage of its power generation, which reduces its energy cost, such that, the prosumer schedules its electric appliances during its preferred time. On the other hand, c-player has less comfort due to the absence of the power generation, such that, the consumer schedules its electric appliances out of its preferred time.
As expected, \lq Ref-sce\rq ~has the highest power consumption. It is observed during peak hours from 18:00 to 20:00. 
\lq Ref-sce\rq ~has also the highest daily energy cost of electricity and with discomfort level equal to zero. 
A significant decrease in the power consumption is observed in the \lq Cost-sce\rq ~from 18:00 to 20:00.

It also achieves the lowest daily energy cost with a reduction of about 40\% compared with the \lq Ref-sce\rq.  However, discomfort level increased by at most 30\%. 
The \lq Cost-discomfort-sce\rq ~has a low power consumption  where the reduction in the daily energy cost is about 37\% compared with the \lq Ref-sce\rq. However, the daily energy cost increased by at most 3\% compared with the \lq Cost-sce\rq.
The discomfort level is maintained in an acceptable level, which is increased by 20\% compared with  the \lq Ref-sce\rq ~and decreased about 10\% compared with the \lq Cost-sce\rq.

In summary, the proposed task scheduling strategy achieved its main objective which is a tradeoff between shifting the  power consumption to time-slots where the daily energy cost is cheaper and maintaining the discomfort of consumers in an acceptable level, thanks to the proposed multiobjective game formulation which optimizes these two conflicting objectives. In contrast with the existing approaches which focus only on the energy cost and impose a direct control on consumers appliances ignoring their comfort, the proposed energy management game achieves a significant results in term of cost as well as consumer's discomfort.

\section{\uppercase{Conclusion}}
\label{sec:conclusion}
In this paper, a non cooperative game theoretic approach has been proposed to implement a DR energy management strategy for competitive residential consumers to obtain a minimum daily energy cost and minimum discomfort level. 
The method presented has been solved using  NSGA-II algorithm. 
In the proposed game theoretic approach, the consumers  play a key role in the energy management game through exploiting  DERs and scheduling their electric appliances locally, in contrast to centralized energy management systems that use DLC approaches that directly control the loads and  impose load shedding to the consumers. 
The developed  energy management game can be used as a useful tool for evaluating the electricity market and also for analyzing the strategic behavior of consumers in competitive electricity markets. 
As a perspective, we will consider additional choices of game theoretic approaches for DSM and more comprehensive decision-making models for consumers based on behavioral sciences.

\bibliography{mybib}

\begin{thebibliography}{20}
\providecommand{\natexlab}[1]{#1}
\providecommand{\url}[1]{\texttt{#1}}
\providecommand{\urlprefix}{URL }
\expandafter\ifx\csname urlstyle\endcsname\relax
  \providecommand{\doi}[1]{doi:\discretionary{}{}{}#1}\else
  \providecommand{\doi}{doi:\discretionary{}{}{}\begingroup
  \urlstyle{rm}\Url}\fi

\bibitem[{Abidi et~al.(2017)Abidi, Smida, Khalgui, Li, and Wu}]{abidi2017multi}
Abidi M.G.; Smida M.B.; Khalgui M.; Li Z.; and Wu N., 2017.
\newblock \emph{Multi-agent oriented solution for forecasting-based control
  strategy with load priority of microgrids in an island mode--Case study:
  Tunisian petroleum platform}.
\newblock \emph{Electric Power Systems Research}, 152, 411--423.

\bibitem[{Al~Zahr et~al.(2017)Al~Zahr, Doumith, and Forestier}]{al2017advanced}
Al~Zahr S.; Doumith E.A.; and Forestier P., 2017.
\newblock \emph{Advanced Demand Response Considering Modular and Deferrable
  Loads under Time-Variable Rates}.
\newblock In \emph{GLOBECOM 2017-2017 IEEE Global Communications Conference}.
  IEEE, 1--6.

\bibitem[{Deb et~al.(2002)Deb, Pratap, Agarwal, and Meyarivan}]{deb2002fast}
Deb K.; Pratap A.; Agarwal S.; and Meyarivan T., 2002.
\newblock \emph{A Fast and Elitist Multiobjective Genetic Algorithm: NSGA-II}.
\newblock \emph{IEEE transactions on evolutionary computation}, 6, no.~2,
  182--197.

\bibitem[{Deng et~al.(2014)Deng, Yang, Chen, Asr, and
  Chow}]{deng2014residential}
Deng R.; Yang Z.; Chen J.; Asr N.R.; and Chow M.Y., 2014.
\newblock \emph{Residential Energy Consumption Scheduling: A Coupled-onstraint
  Game Approach}.
\newblock \emph{IEEE Transactions on Smart Grid}, 5, no.~3, 1340--1350.

\bibitem[{Eksin et~al.(2015)Eksin, Deli{\c{c}}, and Ribeiro}]{eksin2015demand}
Eksin C.; Deli{\c{c}} H.; and Ribeiro A., 2015.
\newblock \emph{Demand Response Management in Smart Grids with Heterogeneous
  Consumer Preferences}.
\newblock \emph{IEEE Transactions on Smart Grid}, 6, no.~6, 3082--3094.

\bibitem[{Fadel et~al.(2015)Fadel, Gungor, Nassef, Akkari, Malik, Almasri, and
  Akyildiz}]{fadel2015survey}
Fadel E.; Gungor V.C.; Nassef L.; Akkari N.; Malik M.A.; Almasri S.; and
  Akyildiz I.F., 2015.
\newblock \emph{A survey on wireless sensor networks for smart grid}.
\newblock \emph{Computer Communications}, 71, 22--33.

\bibitem[{Fudenberg and Tirole(1991)}]{fudenberg1991game}
Fudenberg D. and Tirole J., 1991.
\newblock \emph{Game theory, 1991}.
\newblock \emph{Cambridge, Massachusetts}, 393, no.~12, 80.

\bibitem[{Kaddah et~al.(2014)Kaddah, Kofman, and Pioro}]{kaddah2014advanced}
Kaddah R.; Kofman D.; and Pioro M., 2014.
\newblock \emph{Advanced Demand Response Solutions Based on Fine-grained Load
  Control}.
\newblock In \emph{Intelligent Energy Systems (IWIES), 2014 IEEE International
  Workshop on}. IEEE, 38--45.

\bibitem[{Kim et~al.(2013)Kim, Ren, Van Der~Schaar, and
  Lee}]{kim2013bidirectional}
Kim B.G.; Ren S.; Van Der~Schaar M.; and Lee J.W., 2013.
\newblock \emph{Bidirectional energy trading and residential load scheduling
  with electric vehicles in the smart grid}.
\newblock \emph{IEEE Journal on Selected Areas in Communications}, 31, no.~7,
  1219--1234.

\bibitem[{Li et~al.(2011)Li, Chen, and Low}]{li2011optimal}
Li N.; Chen L.; and Low S.H., 2011.
\newblock \emph{Optimal demand response based on utility maximization in power
  networks}.
\newblock In \emph{Power and Energy Society General Meeting, 2011 IEEE}. IEEE,
  1--8.

\bibitem[{Ma et~al.(2013)Ma, Alkadi, Cappers, Denholm, Dudley, Goli, Hummon,
  Kiliccote, MacDonald, Matson et~al.}]{ma2013demand}
Ma O.; Alkadi N.; Cappers P.; Denholm P.; Dudley J.; Goli S.; Hummon M.;
  Kiliccote S.; MacDonald J.; Matson N.; et~al., 2013.
\newblock \emph{Demand Response for Ancillary Services}.
\newblock \emph{IEEE Transactions on Smart Grid}, 4, no.~4, 1988--1995.

\bibitem[{Meskina et~al.(2017)Meskina, Doggaz, Khalgui, and
  Li}]{meskina2017multiagent}
Meskina S.B.; Doggaz N.; Khalgui M.; and Li Z., 2017.
\newblock \emph{Multiagent framework for smart grids recovery}.
\newblock \emph{IEEE Transactions on Systems, Man, and Cybernetics: Systems},
  47, no.~7, 1284--1300.

\bibitem[{Mosbahi and Khalgui(2016)}]{mosbahi2016new}
Mosbahi O. and Khalgui M., 2016.
\newblock \emph{New solutions for optimal power production, distribution and
  consumption in smart grids}.
\newblock \emph{International Journal of Modelling, Identification and
  Control}, 26, no.~2, 110--129.

\bibitem[{Nash(1951)}]{nash1951non}
Nash J., 1951.
\newblock \emph{Non-cooperative games}.
\newblock \emph{Annals of mathematics}, 286--295.

\bibitem[{Ning et~al.(2017)Ning, Tang, Chen, Wang, Zhou, and Gao}]{ning2017bi}
Ning J.; Tang Y.; Chen Q.; Wang J.; Zhou J.; and Gao B., 2017.
\newblock \emph{A bi-level coordinated optimization strategy for smart
  appliances considering online demand response potential}.
\newblock \emph{Energies}, 10, no.~4, 525.

\bibitem[{Safamehr and Rahimi-Kian(2015)}]{safamehr2015cost}
Safamehr H. and Rahimi-Kian A., 2015.
\newblock \emph{A cost-efficient and reliable energy management of a micro-grid
  using intelligent demand-response program}.
\newblock \emph{Energy}, 91, 283--293.

\bibitem[{Salinas et~al.(2013)Salinas, Li, and Li}]{salinas2013multi}
Salinas S.; Li M.; and Li P., 2013.
\newblock \emph{Multi-objective optimal energy consumption scheduling in smart
  grids}.
\newblock \emph{IEEE Transactions on Smart Grid}, 4, no.~1, 341--348.

\bibitem[{Samadi et~al.(2012)Samadi, Mohsenian-Rad, Schober, and
  Wong}]{samadi2012advanced}
Samadi P.; Mohsenian-Rad H.; Schober R.; and Wong V.W., 2012.
\newblock \emph{Advanced demand side management for the future smart grid using
  mechanism design}.
\newblock \emph{IEEE Transactions on Smart Grid}, 3, no.~3, 1170--1180.

\bibitem[{Vivekananthan et~al.(2014)Vivekananthan, Mishra, Ledwich, and
  Li}]{vivekananthan2014demand}
Vivekananthan C.; Mishra Y.; Ledwich G.; and Li F., 2014.
\newblock \emph{Demand Response for Residential Appliances Via Customer Reward
  Scheme}.
\newblock \emph{IEEE transactions on smart grid}, 5, no.~2, 809--820.

\bibitem[{Yang et~al.(2013)Yang, Tang, and Nehorai}]{yang2013game}
Yang P.; Tang G.; and Nehorai A., 2013.
\newblock \emph{A game-theoretic approach for optimal time-of-use electricity
  pricing}.
\newblock \emph{IEEE Transactions on Power Systems}, 28, no.~2, 884--892.

\end{thebibliography}

\end{document}